\documentstyle[epsfig,onecolumn]{mn}   % For submission
\newif\ifAMStwofonts
\newcommand{\ub}{$U\!-\!B\,$}
\newcommand{\bv}{$B\!-\!V\,$}
\newcommand{\vi}{$V\!-\!I\,$}

\newcommand{\uvv}{$m(1500)\!-\!V\,$}
\ifoldfss
  \ifCUPmtlplainloaded \else
    \NewTextAlphabet{textbfit} {cmbxti10} {}
    \NewTextAlphabet{textbfss} {cmssbx10} {}
    \NewMathAlphabet{mathbfit} {cmbxti10} {} % for math mode
    \NewMathAlphabet{mathbfss} {cmssbx10} {} %  "   "    "
  \fi
  \ifAMStwofonts
    \ifCUPmtlplainloaded \else
      \NewSymbolFont{upmath} {eurm10}
      \NewSymbolFont{AMSa} {msam10}
      \NewMathSymbol{\upi}     {0}{upmath}{19}
      \NewMathSymbol{\umu}     {0}{upmath}{16}
      \NewMathSymbol{\upartial}{0}{upmath}{40}
      \NewMathSymbol{\leqslant}{3}{AMSa}{36}
      \NewMathSymbol{\geqslant}{3}{AMSa}{3E}

      \let\leq=\leqslant \let\le=\leqslant
       \let\ge=\geqslant
    \fi
  \fi
\fi % End of OFSS

\ifnfssone
  \newmathalphabet{\mathit}
  \addtoversion{normal}{\mathit}{cmr}{m}{it}
  \addtoversion{bold}{\mathit}{cmr}{bx}{it}
  \newmathalphabet{\mathbfit} % math mode version of \textbfit{..}
  \addtoversion{normal}{\mathbfit}{cmr}{bx}{it}
  \addtoversion{bold}{\mathbfit}{cmr}{bx}{it}
  \newmathalphabet{\mathbfss} % math mode version of \textbfss{..}
  \addtoversion{normal}{\mathbfss}{cmss}{bx}{n}
  \addtoversion{bold}{\mathbfss}{cmss}{bx}{n}
  \ifAMStwofonts
    \ifCUPmtlplainloaded \else
      \UseAMStwoboldmath
      \makeatletter
      \new@mathgroup\upmath@group
      \define@mathgroup\mv@normal\upmath@group{eur}{m}{n}
      \define@mathgroup\mv@bold\upmath@group{eur}{b}{n}
      \edef\UPM{\hexnumber\upmath@group}
      \new@mathgroup\amsa@group
      \define@mathgroup\mv@normal\amsa@group{msa}{m}{n}
      \define@mathgroup\mv@bold\amsa@group{msa}{m}{n}
      \edef\AMSa{\hexnumber\amsa@group}
      \makeatother
      \mathchardef\upi="0\UPM19
      \mathchardef\umu="0\UPM16
      \mathchardef\upartial="0\UPM40
      \mathchardef\leqslant="3\AMSa36
      \mathchardef\geqslant="3\AMSa3E

      \let\leq=\leqslant \let\le=\leqslant
       \let\ge=\geqslant
    \fi
  \fi
\fi % End of NFSS release 1

\ifnfsstwo
  \DeclareMathAlphabet{\mathbfit}{OT1}{cmr}{bx}{it}
  \SetMathAlphabet\mathbfit{bold}{OT1}{cmr}{bx}{it}
  \DeclareMathAlphabet{\mathbfss}{OT1}{cmss}{bx}{n}
  \SetMathAlphabet\mathbfss{bold}{OT1}{cmss}{bx}{n}
  \ifAMStwofonts
    \ifCUPmtlplainloaded \else
      \DeclareSymbolFont{UPM}{U}{eur}{m}{n}
      \SetSymbolFont{UPM}{bold}{U}{eur}{b}{n}
      \DeclareSymbolFont{AMSa}{U}{msa}{m}{n}
      \DeclareMathSymbol{\upi}{0}{UPM}{"19}
      \DeclareMathSymbol{\umu}{0}{UPM}{"16}
      \DeclareMathSymbol{\upartial}{0}{UPM}{"40}
      \DeclareMathSymbol{\leqslant}{3}{AMSa}{"36}
      \DeclareMathSymbol{\geqslant}{3}{AMSa}{"3E}

      \let\leq=\leqslant \let\le=\leqslant
       \let\ge=\geqslant
    \fi
  \fi
\fi % End of NFSS release 2

\ifCUPmtlplainloaded \else
  \ifAMStwofonts \else % If no AMS fonts
    \def\upi{\pi}
    \def\umu{\mu}
    \def\upartial{\partial}
  \fi
\fi

\title{Globular Clusters as Probes of Galaxy Evolution: NGC\,5128}
\author[S.K. Yi et al.]
{S.K. Yi,$^{1}$\footnote{yi@astro.ox.ac.uk} 
E. Peng,$^{2,3}$ H. Ford,$^{2,4}$ S. Kaviraj$^1$ and S.-J. Yoon$^{1,5}$\\
$^1$University of Oxford, Astrophysics, Keble Road, Oxford OX1 3RH\\
$^2$Department of Physics and Astronomy, Johns Hopkins University, 3400 North Charles Street, Baltimore, MD 21218, USA\\
$^3$Department of Physics and Astronomy, Rutgers University, 136 Frelinghuysen Road, Piscataway, NJ, 08854, USA\\
$^4$Space Telescope Science Institute, 3700 San Martin Drive, Baltimore, MD, 21219, USA\\
$^5$Center for Space Astrophysics, Yonsei University, Seoul 120-749, South Korea}
\date{Draft \today}

\pagerange{\pageref{firstpage}--\pageref{lastpage}}
\pubyear{2003}

\begin{document}

\maketitle

\label{firstpage}

\begin{abstract}

We present the results of our analysis of the photometric
data of globular clusters in the elliptical galaxy NGC\,5128 (Cen A).
We show that the integrated colour \ub\ can be an effective metallicity 
indicator for $simple$ stellar populations. This is because it is 
sensitive to metallicity via the opacity effect but relatively insensitive
to the effective 
%main-sequence temperature 
main sequence turn-off temperature 
of the population (and thus to age)
when $T_{\rm eff} \approx 7000$ -- 12000\,K, that is, when the Balmer
Jump is strong. 
This flat \ub\ vs $T_{\rm eff}$ relation is a result of the fact that 
the blueing effect of the optical continuum with increasing temperature
is temporarily stopped by the Balmer Jump which becomes greater with 
temperature in this range.
In this study we use \ub\ and \bv\ as metallicity and
age indicators, respectively. We first show that the use of the 
\ub\ vs \bv\ two-colour diagram roughly yields the metallicities 
and ages of the Milky Way globular clusters independently determined,
and then apply the technique to the clusters in NGC\,5128. 
There is a large range in \ub, which corresponds to
[Fe/H] of $-$2.0 through over $+$0.3. 
The large
uncertainties from the data and the models currently prevent us from pinning
down their ages and metallicities. 
%Yet, for a given \ub, 
%the scatter in \bv\ is very small, potentially indicating an intrinsically
%narrow age-metallicity relation. The derived mean age estimate seems 
%to gradually decrease with increasing metallicity.
Although a constant age for all these clusters cannot be ruled out,
there is a hint of the metal-rich clusters being younger.
Significance of these results and caveats of the analysis are discussed.

\end{abstract}


\section{Introduction}

When and how big elliptical galaxies developed their present
shape in terms of mass and light are among the most debated questions
in astrophysics.
The classical view assumes that most elliptical galaxies
formed at high redshifts in a monolithic way via a very efficient
starburst (Larson 1974), while the more recent, hierarchical galaxy
formation theory suggests that they are more recent products of mergers of
older disk galaxies (Toomre 1977; Kauffman 1996).
Both scenarios can account for many important
observational phenomena, but their implications for the formation
and morphological evolution of galaxies are significantly different.
Consequently, numerous theoretical and observational investigations aim
to find a more realistic scenario.

Galaxy integrated colours are among the most popular tools to study this
issue. Changes in rest-frame optical colours are often interpreted as 
evidence of ``passive'' evolution of early-type galaxies since
high redshifts (Bower, Lucy, \& Ellis 1992).
However, the actual optical colour evolution of
old populations is so small that the measured colour evolution can be
similarly well reproduced by different theoretical
pictures (Kauffman 1996).  Yi (2003a) points out that the change of optical
colours as a function of time becomes hardly noticeable after the age 
6 -- 7~Gyr.
This means that if the mean age of the stars in an early-type galaxy is
greater than 6 -- 7~Gyr, it is extremely difficult to derive the stellar ages
based on integrated colours and thus to pin down the correct galaxy 
evolution scenario.
Besides, both of the two competing views suggest mean stellar ages
as large as 7~Gyr anyway if the galaxies are in
large dark matter halos, although the ranges of stellar ages are
significantly different from one another. In this regard,
it seems hopeless to use the integrated light of galaxies to derive
their star formation and galaxy evolution history.
Spatially resolved stars seem to be required for the task!

Though not quite spatially resolved, globular clusters (GCs) provide 
excellent alternatives. They are just bright enough for easy detection.
The stars in individual clusters are more or less of the same age and
chemical composition, like our own Milky Way globular clusters
(MWGCs), and thus easy to model. Accurate measurements of their
ages and metallicities will give us important hints to the
formation and evolution of the host galaxy. 
A significant hindrance of GC studies is, however, that their
integrated colours are often degenerate in terms of age and metallicity,
and thus even having multiple colour information is not sufficient to 
accurately determine their ages. 
The so-called age-metallicity degeneracy (hereafter
AMD) was first systematically pointed out by Worthey (1994).

Lately, Yi (2003b) claimed that the use of short-wavelength
broad-band colours (such as \ub) may break the AMD because they are
quite insensitive to age. We present the further details of why
\ub\ behaves that way. We apply the technique to Peng, Ford, \& Freeman's
(2003, submitted) $UBV$-band integrated photometry of 
the GCs in NGC\,5128, the giant elliptical galaxy in the
Centaurus group.

From this analysis we found evidence that may indicate 
(1) a large range of age ($2 \la {\rm Age (Gyr)} \la 11$), 
(2) a large range of metallicity ($-2.0 \la {\rm [Fe/H]} \la +0.3$), 
and (3) a possible age-metallicity relation (AMR). 
This technique is easy to apply to 
extragalactic GC systems and may provide important clues to galaxy 
evolution. We discuss significance, caveats and the future prospects.

\section{Broad-band Colours of Globular Clusters}

Broad-band colours of GCs have been announced useful
for extragalactic evolution studies many times before (to cite
only a few, Crampin \& Hoyle 1961; van den Bergh 1969; Harris \&
Racine 1979; Elson \& Walterbos 1988; Girardi \& Bica 1993; Zepf,
Ashman, \& Geisler 1995; Cohen, Blakeslee, \& Ryzhov 1998).
Globular clusters are generally regarded as ``simple'' stellar
populations, that is, made up of stars of the same age and
chemical composition because stars in each cluster probably formed
out of the same gas cloud nearly simultaneously. This makes them easy
targets of population synthesis models and consequently effective
tracers of galaxy evolution.

\begin{figure}
\centering
\epsfig{figure=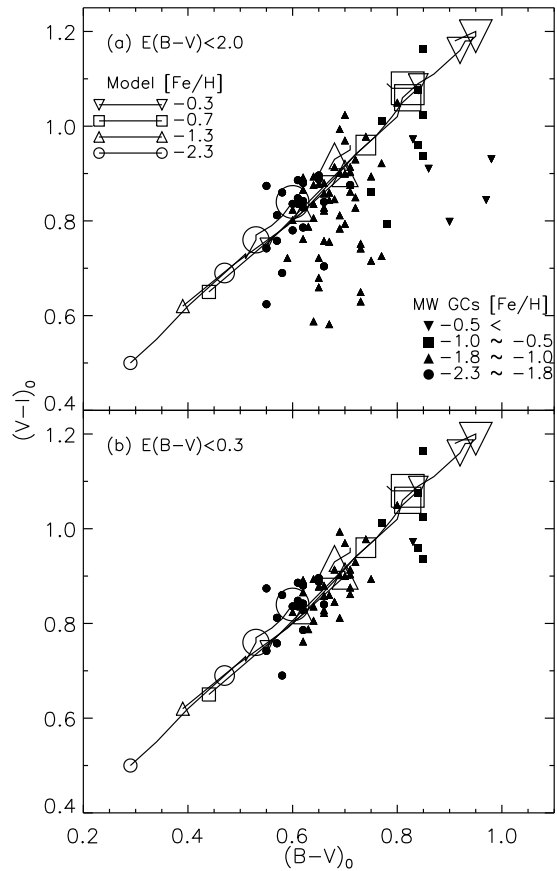,width=0.45\textwidth}
\vspace{0cm} \caption{ Integrated $BVI$ colours of Milky Way
globular clusters (MWGCs) from Harris (1996). The popular $BVI$
colours suffer from the age-metallicity degeneracy (AMD). Data are
compared to population synthesis models of four metallicities (Yi
2003a). Each model sequence covers 1--15\,Gyr, where 1 (smallest),
5, 9, 13 (largest) Gyr models are marked with open symbols. Note
how tight the AMD appears in the low-reddening data ($E(B-V)<0.3$) as
shown in (b). Also notable is that, MWGC colours are reasonably
reproduced by the models of the measured metallicities and of their
isochrone-derived ages. But without knowing their metallicities a
priori, $BVI$ colours alone cannot tell us the ages because of the
AMD. }
\end{figure}

As mentioned in \S1, a serious hindrance is that broad-band
colours are known to suffer from the AMD. Fig.~1 shows a clear
example of the AMD in $BVI$ colours, the most popular colours for
distant stellar population studies. The MWGC data are compared to
the population synthesis models of Yi (2003a; Yi \& Yoon in preparation). 
Table 1 lists the models for select metallicities and ages. 
Their models are for simple stellar populations and are based on
Salpeter initial mass function and the Yale stellar models.
Four models (lines) in Fig.~1 are for four metallicities; the
bluest models with open circles on top being the most metal-poor
of all. Each model (line) is a sequence of age 1--15\,Gyr with 1
(smallest), 5, 9, 13\,Gyr (biggest) models marked with open
symbols. The MWGC colours are based on the same symbol code except
for the most metal-poor bin; the filled circles denote the most
metal-poor MWGCs with the mean metallicity of roughly [Fe/H]=$-$2,
while the most metal-poor models are for [Fe/H]=$-2.3$. Fig.1-(a)
shows the entire MWGC data, while (b) shows only low reddening 
($E(B-V)\leq0.3$) data, from Harris (1996). 
%For \ub\ reddening
%correction, we use $E(U-B) = 1.12 E(B-V)$ from Schlegel,
%Finkbeiner, \& Davis (1998, Landolt calibration in Table~5).
For \ub\ reddening
correction in Fig.2, we use $E(U-B) = 1.12 E(B-V)$ from Schlegel,
Finkbeiner, \& Davis (1998, Landolt calibration in Table~5).

Note firstly in Fig.1-(b) how tight the AMD appears in the $BVI$
colours. Secondly, the MWGC colours are reasonably reproduced by
the models of their spectroscopically measured metallicities and
of their isochrone-derived ages. But the four models are
virtually on the same extended line; and thus neither of these
colours provides age or metallicity information unless one of them
(age or metallicity) is known a priori.

Yi (2003a) recently claimed that the use of far-UV-to-optical
colours (e.g., \uvv) may avoid the AMD and separate old ($\ga
10$~Gyr) populations from young ones. This is because the age
effect in far-UV-to-optical colours is virtually orthogonal to the
metallicity effect $when$ a population is old enough to develop a
blue horizontal-branch. Unfortunately, far-UV data are only
available through space telescopes and thus difficult to obtain.

\begin{figure}
%\centering \epsfig{figure=plot/MW_bv_ub_bw.eps,width=0.45\textwidth}
\centering \epsfig{figure=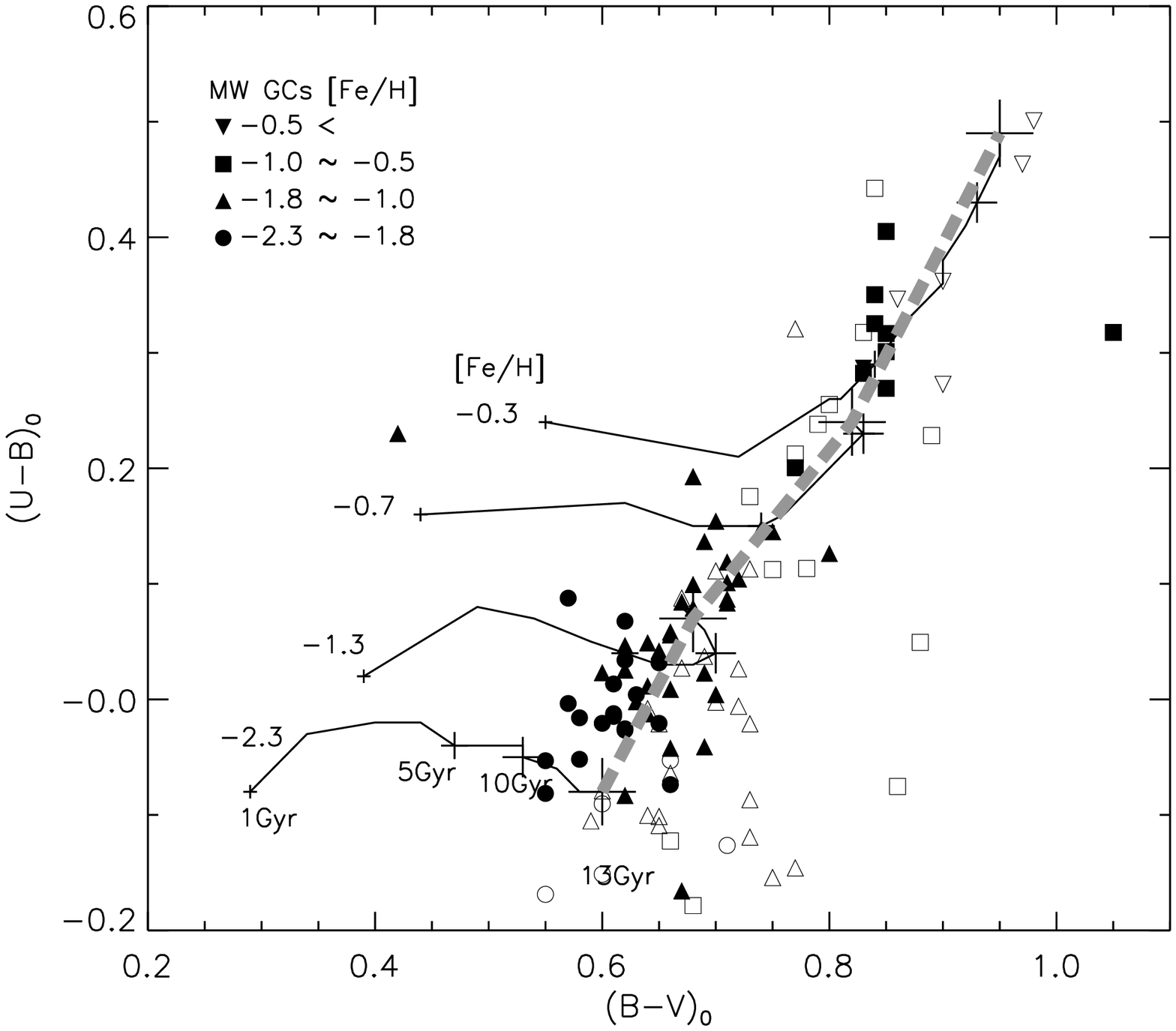,width=0.9\textwidth}
\vspace{0cm} \caption{ Integrated $UBV$ colours of the MWGCs
from Harris (1996). Filled and open symbols are the data of $E(B-V)<0.3$
and $E(B-V)\ge0.3$, respectively. This figure shows that \ub\ can
break the AMD because it acts as a metallicity indicator using the MWGCs.
See Figure 1 caption for data descriptions. The thick grey dashed line
shows the 13~Gyr sequence, which reasonably fits the MWGC data}
\end{figure}

The \ub\ colour is easier to obtain and may
still be capable of breaking the AMD at least partially. Yi (2003b)
showed that \ub\ is sensitive to metallicity but not to age if 
metallicity is low. We compared the colours of the MWGCs to Yi's
population synthesis models to test this. The result is shown in Fig.~2.
Indeed, the four metallicity models (lines) are separated from one
another very clearly. All models span 1 through 13~Gyr, with 1
(smallest), 5, 10, 13~Gyr (largest) models are marked with cross
symbols. Our models reproduce the MWGC data reasonably well in
both colours. A small discrepancy of order 0.03--0.04~mag (in
particular in \ub) may appear noticeable to some readers. Such
offsets are currently unavoidable given the uncertainties in the
$T_{\rm eff}$-colour transformation (Lejeune et al. 1997). The
thick vertical dashed line is the 13~Gyr sequence derived from the four
metallicity models. It appears to be a good representation of the
MWGC data, although the scatter is large.

\begin{figure}
\centering
\epsfig{figure=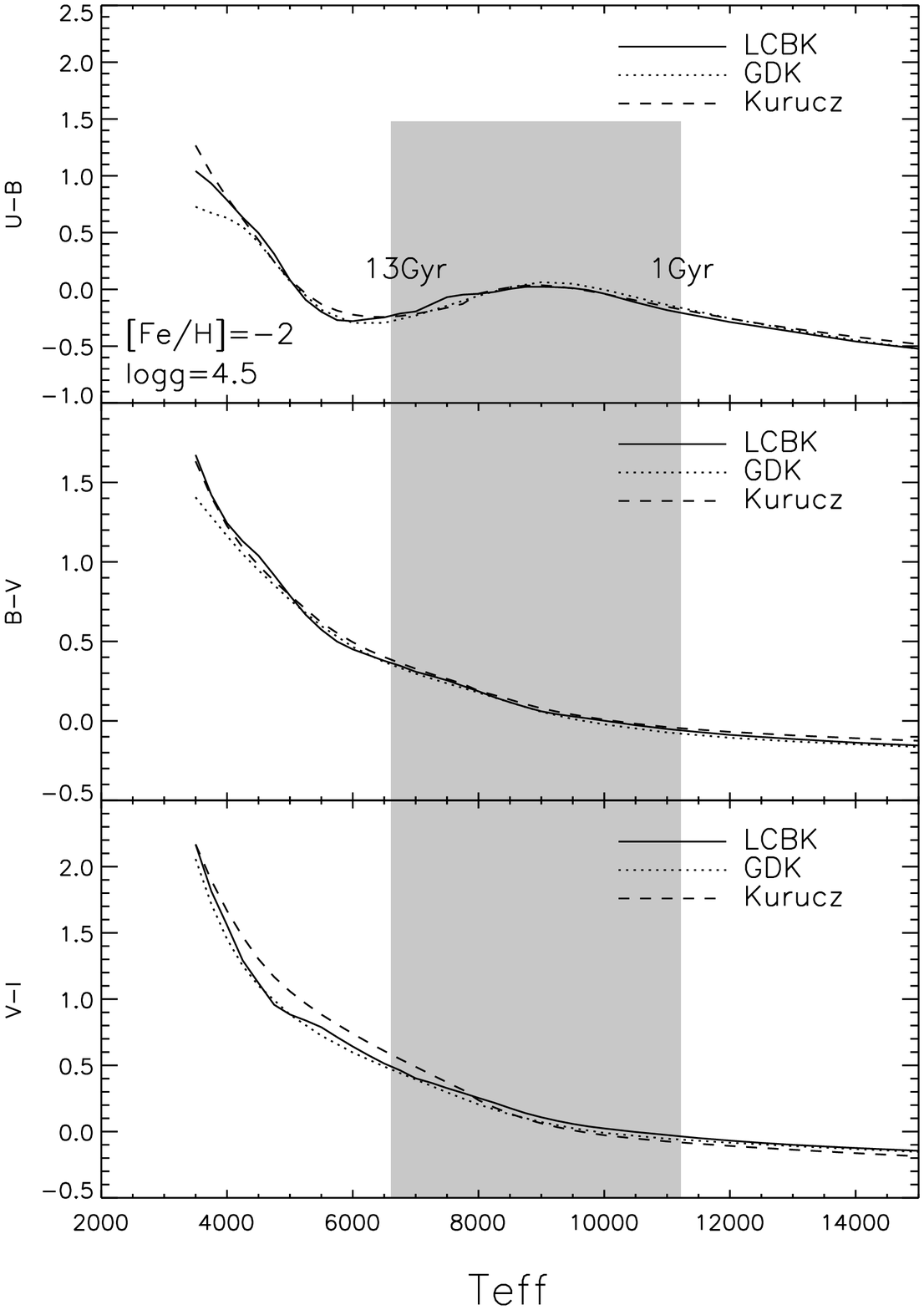,width=0.9\textwidth}
\vspace{0cm} \caption{ $T_{\rm eff}$-colour relations for
metal-poor dwarfs. Unlike \bv\ and \vi, the \ub\ colour shows no
monotonic variation as a function of temperature for the age range
1--13\,Gyr. The shaded area is the temperature range of the MSTO
for the population age of 1--13~Gyr.  
Three $T_{\rm eff}$-colour relations, shown
for comparison, all exhibit the same trend. They are the
Lejeune et al.-calibrated Kurucz calibration
(LCBK), the Green, Demarque, \& King semi-empirical calibration
(Gree, Demarque, \& King 1987), and the Kurucz theoretical 
calibration (Kurucz 1992). Refer to Yi (2003a) for
details about these calibrations. }
\end{figure}

\begin{figure}
\centering
\epsfig{figure=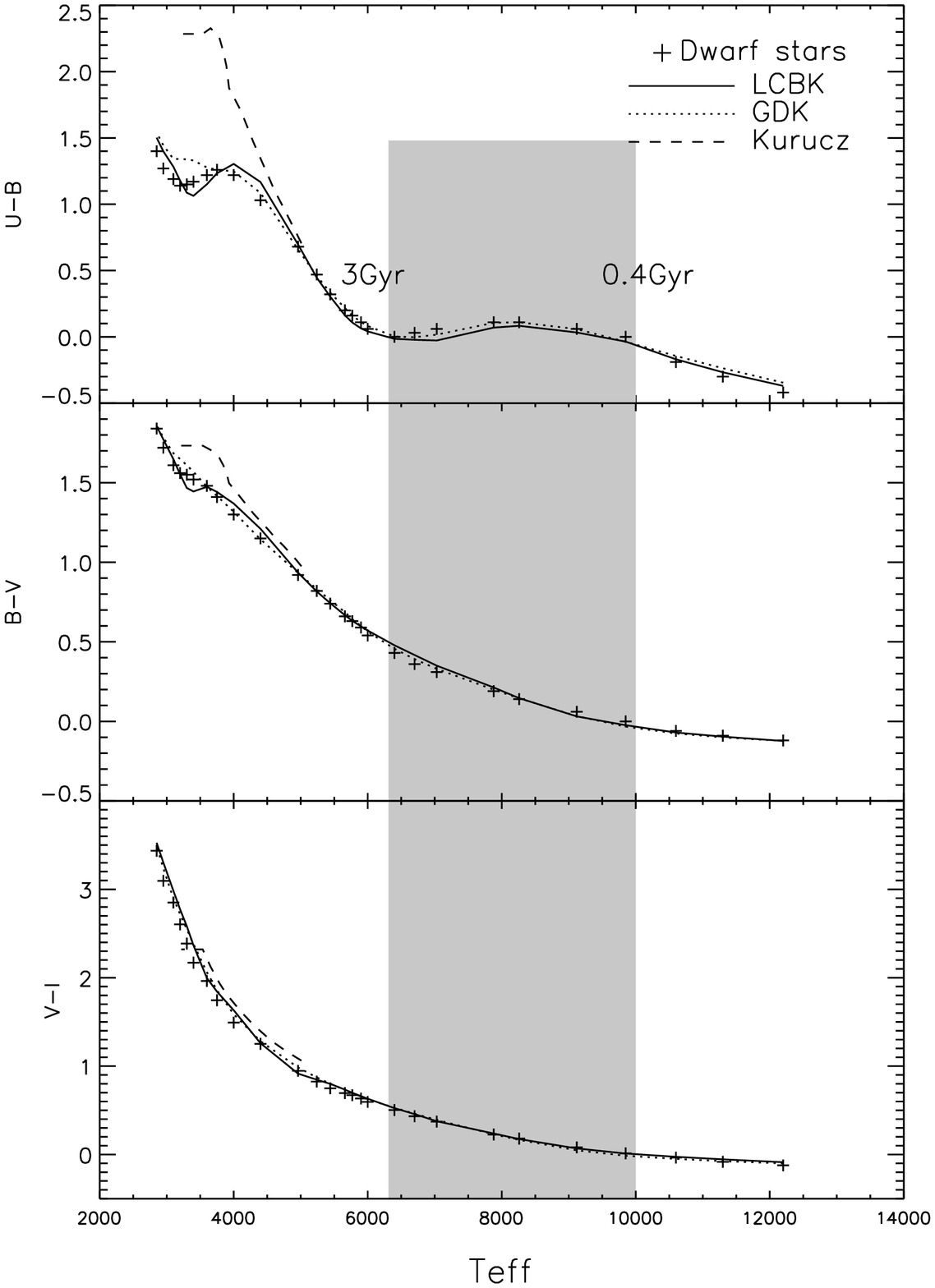,width=0.9\textwidth}
\vspace{0cm} \caption{ Same as Fig.~3 but for the solar
metallicity. Also shown are stellar sample (plus signs) kindly
provided by Worthey (private communication). Note that \ub\ begins
to have a monotonic behaviour at large ages. For this reason, \ub\
serves as a metallicity indicator at smaller ages (0.4--3\,Gyr)
if metallicity is high. }
\end{figure}

Then we wonder what causes \ub\ to be so insensitive to age. For
instance, at low metallicities, \ub\ is nearly flat (within 0.1~mag 
variation) for a good part of the age range (1--15~Gyr when 
[Fe/H]=$-$2.3, 1--9~Gyr when ${\rm [Fe/H]} \approx -0.7$). This is
answered in the next two figures. Figs. 3 and 4 show optical
colours as a function of temperature. The shaded box denotes the
main-sequence turn-off (MSTO) temperature for the age range in question
(1--13~Gyr). Unlike \bv\ or \vi, \ub\ does not exhibit a monotonic
change with respect to temperature. The $S$ shape in the $T_{\rm
eff}$ vs \ub\ relation in Fig.~3 is also visible along the models
in Fig.~2. 
The \ub\ colour is sensitive to metallicity via the opacity effect. 
But it shows a semi-flat curve with effective temperature (and thus with age)
when $T_{\rm eff} \approx 7000$ -- 12000\,K (ages $\approx$ 1--10\,Gyr) 
if [Fe/H]=$-$2 as shown in Fig.~3 and when 
$T_{\rm eff} \approx 6000$ -- 10000\,K (ages 0.4--3\,Gyr) if [Fe/H]=+0.0
as shown in Fig.~4. 
This relatively flat \ub\ vs $T_{\rm eff}$ relation is a result of 
the fact that the blueing effect of the optical continuum with 
increasing temperature is temporarily held back by the Balmer Jump 
which becomes stronger with temperature in this range 
(Binney \& Merrifield 1998).
This is why \ub\ acts as a good metallicity indicator and does so in different
age ranges at different metallicities.

Similar techniques have been used by Rejkuba (2001) and Jord{\' a}n et al.
(2002), who used $U$$-$$V$ and the HST/WFPC2 F336W$-$F547M, respectively, 
as a metallicity indicator. Such colours are substantially better than 
\vi or \bv as a metallicity indicator; but they still change gradually 
with respect to age and thus are not as effective as \ub.

In summary, we have shown that the popular \bv\ and \vi\ colours
suffer from severe cases of the AMD and fail to
serve as age or metallicity indicators. The \ub\ colour, on the
contrary, is predicted to be a good metallicity indicator. 
We demonstrate this using the MWGC data. 
As the derived ages of the MWGCs tell us about how Galactic halo
and thick disk have evolved, we hope to apply \ub\ to
extragalactic GCs and find their metallicities and ages in order
to understand the evolution of their host galaxies.

\begin{table*}
\caption{The colour evolution model grids$^a$.}\label{tbl-1}
\begin{center}
\begin{tabular}{rrrrrr}
\hline \hline
[Fe/H] & Age(Gyr) & $U$$-$$B$ & $B$$-$$V$ & $V$$-$$R$ & $V$$-$$I$ \\
\hline
$-$2.3 &      1 &    -0.08 &     0.29 &     0.23 &     0.50\\
$-$2.3 &      2 &    -0.03 &     0.34 &     0.25 &     0.55\\
$-$2.3 &      3 &    -0.02 &     0.40 &     0.29 &     0.62\\
$-$2.3 &      4 &    -0.02 &     0.44 &     0.31 &     0.66\\
$-$2.3 &      5 &    -0.04 &     0.47 &     0.32 &     0.69\\
$-$2.3 &      6 &    -0.04 &     0.51 &     0.34 &     0.73\\
$-$2.3 &      7 &    -0.04 &     0.51 &     0.34 &     0.72\\
$-$2.3 &      8 &    -0.04 &     0.52 &     0.35 &     0.74\\
$-$2.3 &      9 &    -0.04 &     0.53 &     0.36 &     0.76\\
$-$2.3 &     10 &    -0.05 &     0.53 &     0.36 &     0.77\\
$-$2.3 &     11 &    -0.06 &     0.56 &     0.38 &     0.79\\
$-$2.3 &     12 &    -0.08 &     0.58 &     0.39 &     0.81\\
$-$2.3 &     13 &    -0.08 &     0.60 &     0.40 &     0.84\\
$-$2.3 &     14 &    -0.09 &     0.60 &     0.40 &     0.83\\
$-$2.3 &     15 &    -0.09 &     0.62 &     0.41 &     0.85\\
\hline
\end{tabular}
\end{center}
$^a${Table 1 is available in its entirety in the electronic edition of
the Monthly Notices of the Royal Astronomical Society.}\\
\end{table*}

\section{Globular Clusters of NGC\,5128}

NGC\,5128 (Cen A) is the closest giant elliptical galaxy
at distance 3.5--4\,Mpc. 
Owing to its proximity, its GCs have been found early by
Graham \& Phillips (1980) and subsequently studied by many groups
(to cite only a few, Frogel 1984; Jablonka et al. 1996; Minniti et
al. 1996; Rejkuba 2001). Peng et al. (2003) have lately
acquired the $UBV$ data on its 210 GCs.  These observations were taken
with the CTIO 4-meter telescope with its Mosaic imaging camera in June,
2000.  The photometry was calibrated using standard stars from
Landolt (1992).  Descriptions of these observations are given in Peng et
al.\ (2002, 2003)  The catalog we use is of the 210
spectroscopically confirmed GCs.  
While it is always important that GCs have radial velocities consistent
with that of the host galaxy, the high foreground contamination in the
direction of NGC 5128 makes radial velocity selection critically important
for defining a sample of GCs.
%Because of the high level of foreground star
%contamination, it is important that each GC has a radial
%velocity consistent with being part of the NGC\,5128 system.  
All GCs in
this catalog have $250 < v_{helio} < 1000$~km~s$^{-1}$, and it 
is on this data set that we apply the two-colour diagram technique.

Fig.~5 shows the NGC\,5128 GC data compared to our population
models for six metallicities. Data are divided into two groups
depending on the $U$-band measurement errors. The small-error
sample ($U$-band error smaller than 0.2~mag) appear as filled
circles while the large-error data appear as open circles. The
mean error in the whole sample is 0.2\,mag, while that
of the low-error data alone is 0.1\,mag. The mean
errors in colours for the small-error sample are shown as cross 
in the top left.
The arrow in the lower right corner illustrates the Galactic 
extinction, $E(B-V)=0.11$\,mag, estimated using the
Schlegel et al. (1998) formula. The internal reddening in
NGC\,5128 does not seem to add much to this (Frogel 1984; Jablonka
et al. 1996; Rejkuba et al. 2002) unless objects are near its dust
lane. The GCs observed by Peng et al.\ are mostly outside
1~effective radius (approximately 5 arcmin) and away from the dust
lane. Also shown are the low-reddening MWGCs from Fig.~1-(b).

Several features are apparent. Most notable is a large range in
\ub\ in the NGC\,5128 GC sample compared to the MWGCs. This
indicates a wide range of metallicity in the NGC\,5128 sample. 
%Based on the basic principles of
%galactic chemical evolution, this implies that a much greater
%amount of gas was involved in the starburst that formed these
%clusters than that involved in the formation of the MW halo. It
%should be interesting to find the mass of gas that is required to
%produce such a high metallicity in a Galactic halo-type 
%starburst and to compare it to the MW total and halo masses. 
%Such an investigation should consider various critical elements
%in the chemical evolution theory, such as environments, gas inflows
%and outflows, star formation timescales, multiple starbursts, etc.
It is also interesting to notice that the fraction of
extremely metal-poor (${\rm [Fe/H]} \la -1.3$) GCs seems much
smaller in NGC\,5128 than in the MW, which confirms the earlier
result of Zepf \& Ashman (1993) based on the Washington photometry
$C-T_{1}$ colour. Fig.\,6 shows this more clearly.

The rough estimates of metallicity are listed in Table 2. 
Column\,1 lists seven \ub\ bins.
Column\,2 lists the (\ub)-based mean metallicity estimates of the clusters. 
%The estimates are roughly derived from visual inspections of Fig.\,6.
Crude age and metallicity 
estimates were obtained from visual inspection of Figure 6 and 7.
The seven data points with error bars in Fig.\,6 are the mean values  
with standard deviations for the \ub\ bins.
The uncertainty is large due to the large measurement error in \ub\ 
(0.2~mag) in the first place; and it is worse at higher metallicities 
because at high metallicities even the \ub\ vs \bv diagram suffers 
from the AMD and the population synthesis models are poorly tested and
calibrated. 
Column 3 shows the ages again roughly derived from the mean 
values of \bv\ in the \ub\ bins using Fig.\,7. 
The errors are translated simply from the standard deviation in \bv\ 
assuming that we have no prior information on metallicity. 
For the upper error estimation, we assume the maximum age of 14\,Gyr.
%For a given value of \ub\ (though with large errors), the scatter
%in \bv\ is small, which can be interpreted as evidence of a narrow 
%range in age for a given metallicity group.
For a given value of \ub\ (though with large errors), the scatter
in \bv\ is small, even compared to the Milky Way GC data.
If we can reduce the measurement errors, we would be able to pin down
their ages.
Figure 7 also shows the MWGC data (with dashed error bars)
determined by the same \ub\ scheme (but only for the first five 
\ub\ bins). A total of 60 MWGCs are used for this measurement,
and 18, 22, 9, 4, and 5 are in each of the five bins from blue
to red. We excluded NGC\,7492 data point in computing the mean points
because it has an abnormally
blue \bv\, colour of 0.42 according to the Harris table, which is not 
easy to reconcile with other cluster colours if they belong to the same
age group. Because of the small number statistics, only the first
two points are significant. Besides, redder GCs on the 
average suffer from a higher extinction and thus less reliable.
Keeping these in mind, the five data points are still consistent
with a single-age hypothesis in the range 12--14\,Gyr.

\begin{figure}
\centering
\epsfig{figure=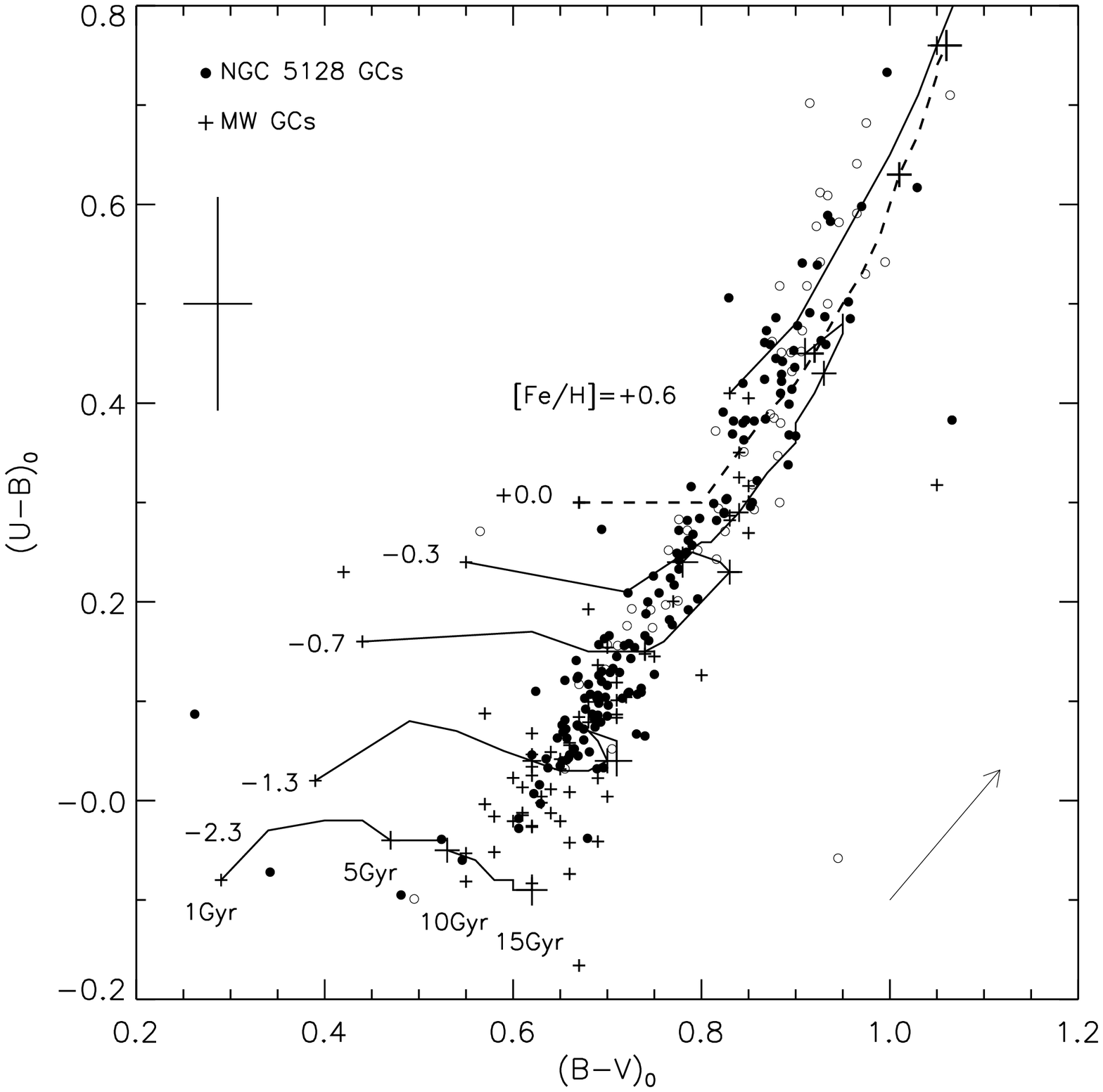,width=0.9\textwidth}
\vspace{0cm} \caption{ Same as Fig.~2, but for NGC\,5128 GCs 
from Peng et al. (2003). The MWGCs are the low-reddening data of Fig.~1-(b).
The large-error GCs  ($U$-band error $> 0.2$~mag) and the small-error GCs 
($U$-band error $\le 0.2$~mag) are shown as open and filled circles,
respectively. The typical errors of filled circles are shown as a large cross 
in the top left. The arrow in the lower-right corner
illustrates the estimated Galactic extinction towards NGC\,5128: 0.11\,mag. 
We ignore internal reddening in NGC\,5128.}
\end{figure}

The scatter of the data in this diagram is strikingly 
small compared to those in the MWGC data. 
The widths in the \ub\ and \bv\ scatters are only of similar
scale to the measurement errors; and, thus, much of the scatters probably
come from the observational errors. 
%This implies that the $sequence$ of
%the NGC\,5128 GC data in this diagram is intrinsically narrow, 
%indicating narrow ranges of age and metallicity $along$ the sequence.
%Such a sequence may lead to an age-metallicity relation (AMR). 
This implies that the $sequence$ of
the NGC\,5128 GC data in this diagram is intrinsically narrow.
Such a sequence, if it can be further narrowed down by achieving higher 
quality data, might lead to an age-metallicity relation (AMR).
For example, the most metal-poor GCs seem similar in colours to their
MW counterparts, suggesting they are comparable in age; but metal-rich
GCs are more consistent with younger models. Fig.~7 shows this
more clearly. Indeed, redder GCs appear younger.
The bluest, and most metal-poor, GCs ($U-B \la 0.1$) seem
as old as $11^{+3}_{-6}$~Gyr. As metallicity (based on \ub) increases
the mean GC age seems to decrease gradually. At \ub\ = 0.3 the
estimated age on the sequence is approximately $3^{+2}_{-1}$~Gyr. 
%If this is real, the data is indicating a dramatic AMR.
If this is real, the data is indicating an AMR.
Such an AMR, if true, would provide an important clue to the star 
formation history of the galaxy. The anti-correlation between age 
and metallicity makes sense from the views of galactic chemical evolution 
theory. 
It should be noted again, however, that the hypothecal AMR is
not significant enough yet due to the uncertainties discussed above.

\begin{figure}
\centering
\epsfig{figure=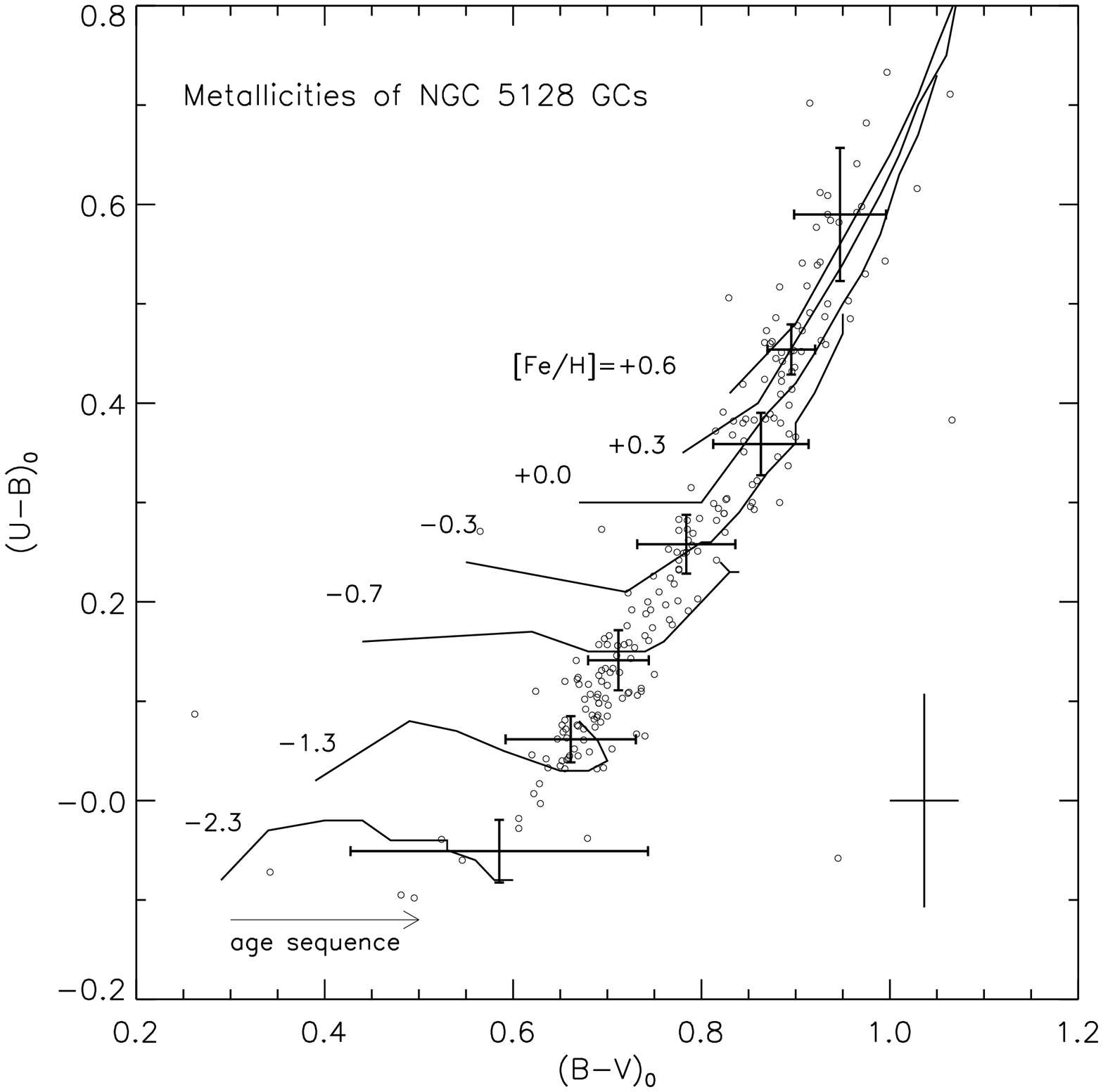,width=0.9\textwidth}
\vspace{0cm} \caption{The age sequences from population synthesis models
for fixed metallicities are compared to the whole data of NGC\,5128 GCs. 
The seven points with error bars along the observed data sequence
are the mean values in the \ub\ bins listed in Table~2.
Metallicity can be estimated with appreciable accuracy 
using \ub. A large range of metallicity is visible.}
\end{figure}

Our last focus is on whether the data show any sign of multiple
metallicity (or age) peaks. Zepf \& Ashman (1993) found two metallicity peaks
in the sample of 61 GCs of NGC\,5128 at ${\rm [Fe/H]} \approx -$0.85 
and $-$0.35 based on $C-T_{1}$. We find similar clusterings 
at two metallicities near $-$1 and 0, as shown in Fig.\,8. 
The separation between the two metallicity peaks seems larger
in our study than that of Zepf \& Ashman (1993). 
Rejkuba (2001) also found two peaks in her 71 GC sample based on 
$U$$-$$V$, but at substantially lower metallicities of 
${\rm [Fe/H]} \approx -$1.7 and $-$0.6. 
In our analysis, we do not find many GCs below 
${\rm [Fe/H]} \approx -$1.3, indicating a notable offset between
her results and ours. The two colours, $U$$-$$V$ and \ub\, have similar 
effects in determining metallicities, but \ub\ is in general less sensitive 
to age and thus being a better metallicity indicator. 
In Fig.~8 we denote the two peaks with rough estimates of metallicity 
based on \ub. However, note that there are large uncertainties in the
metallicity estimates (see Table 2), in particular at high metallicities
because errors are larger for redder GCs: the mean $U$-band errors for 
\ub$< 0.4$ and \ub$\ge 0.4$ GCs are 0.16 and 0.31, respectively.

We do not provide detailed relations between \ub\,and metallicity 
here because they depend on the ages which vary from one galaxy system
to another. Instead we suggest readers to use the model grids
listed in Table 1 for deriving the metallicities of simple stellar 
populations.

\begin{figure}
%\centering \epsfig{figure=plot/n5128_age_bw.eps,width=0.45\textwidth}
\centering \epsfig{figure=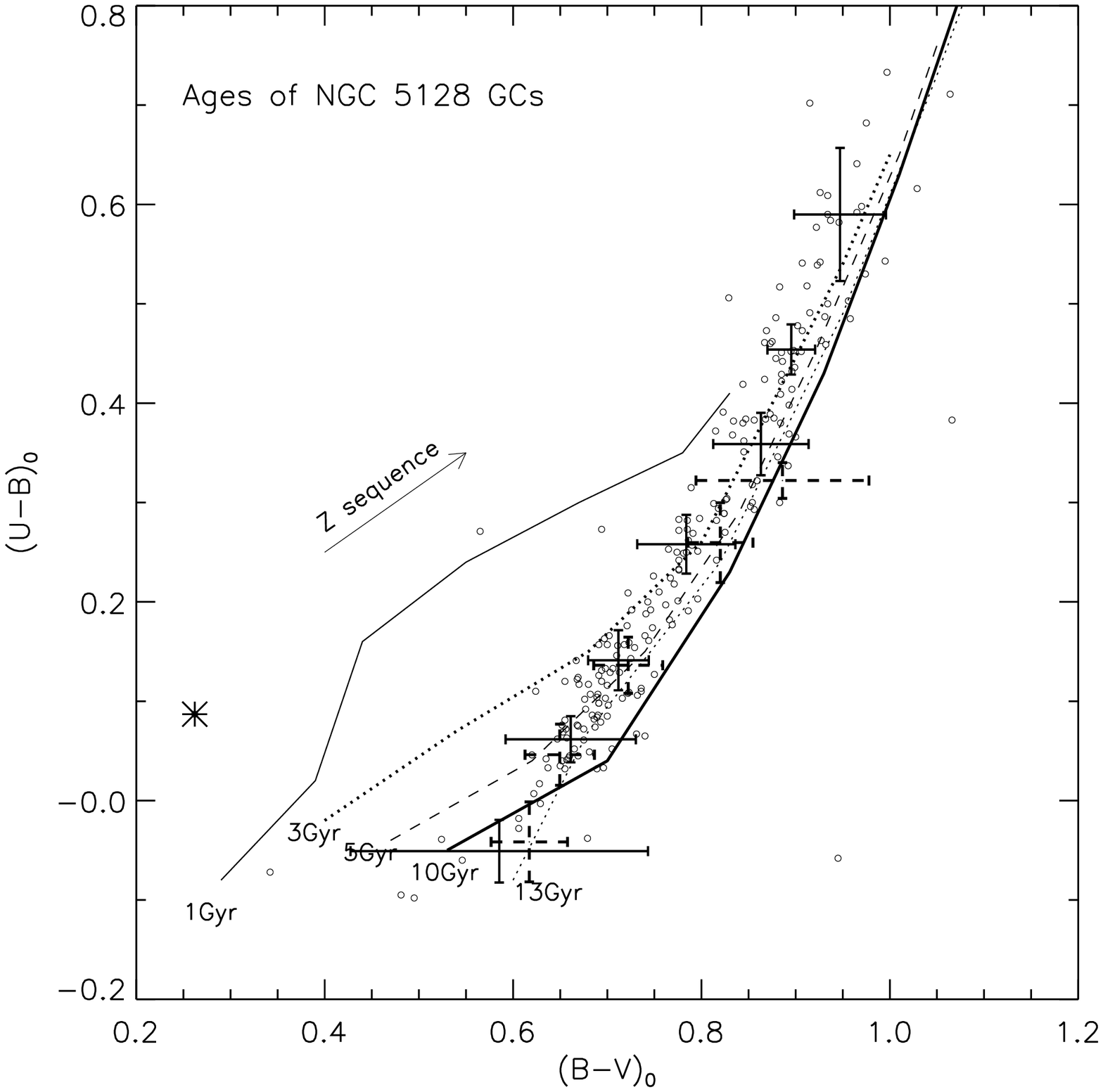,width=0.9\textwidth}
\vspace{0cm} \caption{Same as Fig.~6 but with metallicity sequences
for fixed ages. No single fixed age sequence matches the overall 
colour distribution of the data. An age-metallicity relation,
in the sense of decreasing age with increasing metallicity, is
visible in the data. The bluest GC marked with asterisk is a
young GC candidate.The MWGC data are also shown with dashed-line error bars.
See the text for further details. }
\end{figure}

\begin{table*}
\caption{Age and metallicity derived.}\label{tbl-2}
\begin{center}
\begin{tabular}{cccc}
\hline \hline
\ub\ & [Fe/H]$^a$ & Age(Gyr)$^b$ & N (cluster)\\
\hline
$<0.0$   & $-$2.0$\pm0.3$  & 11$^{+3}_{-6}$ & 10\\
0.0--0.1 & $-$1.1$\pm0.2$  & 7$^{+7}_{-3}$ & 41\\
0.2--0.3 & $-$0.7$\pm0.2$  & 5$^{+9}_{-2}$ & 49 \\
0.3--0.4 & $-$0.3$\pm0.2$  & 3$^{+11}_{-1}$ & 35\\
0.4--0.5 & $-$0.1$\pm0.2$  & 4$^{+10}_{-2}$ & 25\\
0.5--0.6 & +0.3$\pm0.3$  & 3$\pm2$ & 27\\
$0.6<$   & +0.7$\pm0.5$  & 2$^{+12}_{-2}$ & 23\\
\hline
\end{tabular}
\end{center}
$^a${Metallicity derived from \ub.}\\
$^b${Age derived from \bv (assuming the maximum age 14\,Gyr).}\\
\end{table*}

\begin{figure}
\centering
\epsfig{figure=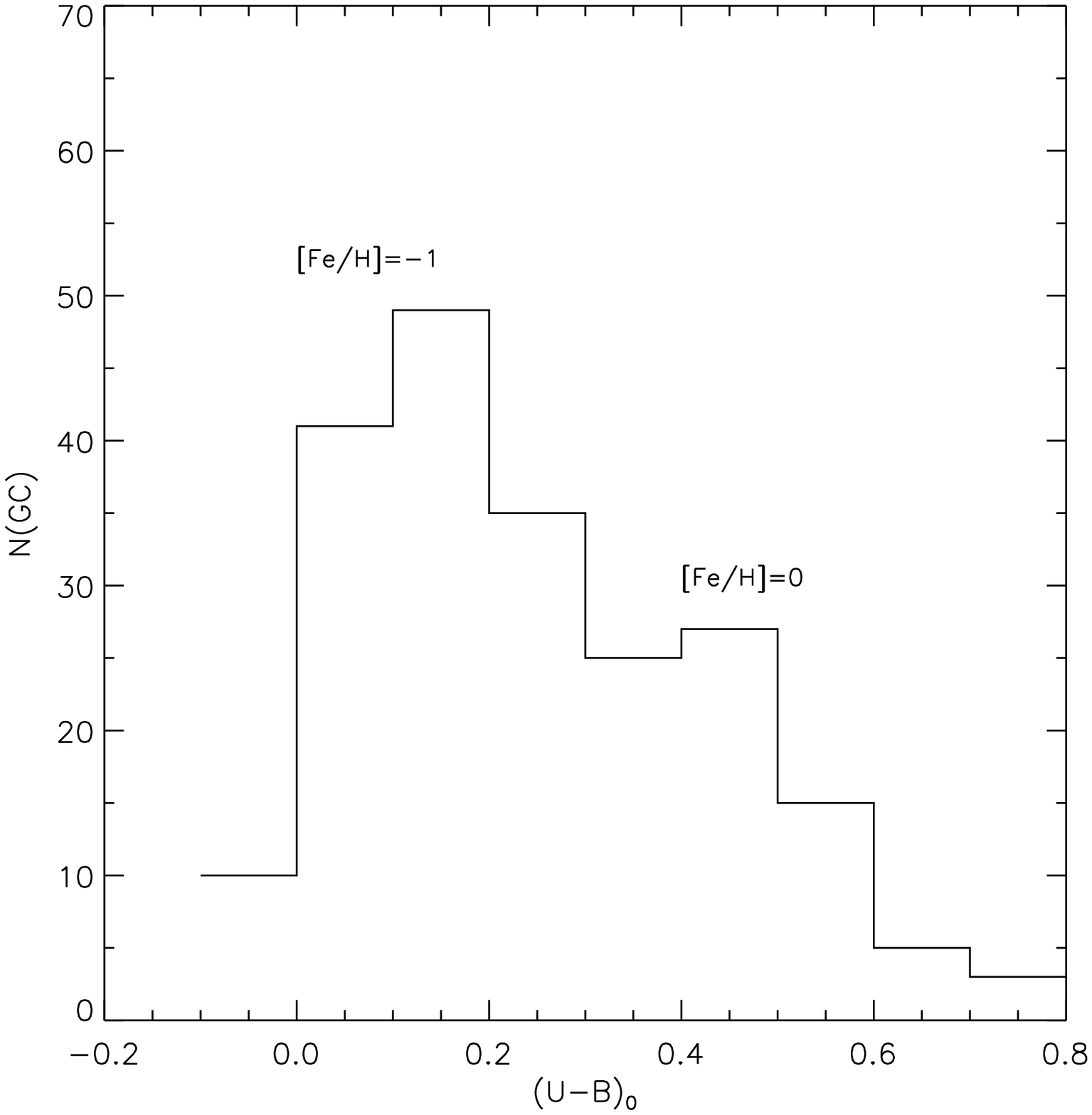,width=0.9\textwidth}
\vspace{0cm} \caption{ The \ub\ colour distribution of NGC\,5128 GCs.
Rough metallicity estimates based on \ub are also marked.}
\end{figure}

\section{Implications on the Evolution of NGC\,5128}

If we take mean values alone from Table~2 (although it is a clear 
oversimplification), our two-colour diagram 
analysis on the NGC\,5128 GCs appears to suggest a large
range in age (2--11~Gyr) and metallicity 
(${\rm [Fe/H]} \approx -2.0$ through +0.3), and possibly an AMR. 
Could this galaxy have been forming star clusters all through this 
age range? Multiple peaks in the colour distributions of 
extragalactic GCs have been interpreted as signs of multiple 
starbursts (e.g., Zepf \& Ashman 1993). Gebhardt \& Kissler-Patig (1999) 
in particular found that half of the galaxies with more than 100 GCs 
observed show \vi\ bimodality and interpreted it as a sign of
double (or more) starbursts.

At this moment, it is worth noting that this galaxy has other 
notable morphological features. 
It shows a prominent dust lane, optical filaments, and many diffuse 
shells (Schiminovich et al. 1994) that are often attributed to a 
recent (100\,Myr ago) merger event. Of the 210 GCs in our sample 
only one at (\bv, \ub)$_{0} \approx$ (0.25, 0.08), marked with asterisk
in Figure 7 seems young enough to have come from such recent merger 
activities. On the other hand, it may be natural not to see many 
young GCs in our sample because our sample avoids the dust-lane 
region, which is the most likely place to find them.

The presence of intermediate-age (a few Gyr) clusters has been 
reported by Minniti et al. (1996). We also find quite a few of them,
which are mostly of a high metallicity. Larsen et al. (2003) have
found such intermediate-age GCs in NGC\,4365 as well.

Gebhardt \& Kissler-Patig (1999) noted that the \vi\ colour bimodality 
is more often found in brighter elliptical galaxies.
Multiple peaks, in particular in age, although not necessarily indicating
multiple major mergers, are compatible with the hierarchical formation
scenario. Kauffmann (1996) suggested that brighter cluster 
elliptical galaxies form their stars at higher redshifts on the average.
NGC\,5128 is a near-$L^{*}$ ($M_{V}^{tot} \approx -21$) $group$ elliptical 
galaxy, and thus its wide-spread starburst signatures, if true, 
would be marginally compatible with hierarchical scenario. 
For a comparison, Cohen et al. (1998) and Jord{\' a}n et al. (2002) suggest
that the colour distributions of GCs of the $cluster$ giant elliptical 
galaxy M\,87 indicate a large constant age of 13--14\,Gyr. Similarly,
Kuntschner et al. (2002) found the GCs in the lenticular galaxy
NGC\,3115 are old and coeval at 11--12\,Gyr.
It will be interesting to obtain $UBV$ data on their GC sample and
apply our two-colour diagram technique.

Because NGC\,5128 is the brightest and probably most massive galaxy in
Centaurus group, it is also plausible that some clusters came from
neighbouring galaxy systems (e.g., C\^{o}t\'{e}, Marzke \& West 1998). 
Such foreign GCs may add to the scatters
in the observed sequence in the two-colour diagram. However, the
present level of the scatters is very small and thus incompatible with
a scenario where many of the GCs formed in independent starbursts that
have distinctively different age-metallicity relations.

\section{Summary and Discussion}

We present the result of our analysis on the $UBV$ multi-band 
photometric data of GCs in the field elliptical galaxy NGC\,5128.
We used \ub\ and \bv\ colours as metallicity and age indicators, respectively.
At low metallicities the \ub\ colour is a good metallicity indicator
because it is sensitive to metallicity but not to age.
We have given a theoretical explanation for these phenomena.
The \ub\ vs \bv\ two-colour diagram technique is able to differentiate
old populations from young ones quite successfully. We first show that 
this technique works very well on the MWGCs; our models match the MWGC
colours at their independently-derived metallicities and ages.

We applied the same technique to the NGC\,5128 GC data  
and found the following. There is a large range 
in \ub, which corresponds to [Fe/H]=$-$2.0 through possibly over +0.3. 
%and a large spread in \bv, which corresponds to age of 2--11\,Gyr.  
For a given \ub, the spread in \bv is very small, making the age
estimation possible. Despite the fact that the observational errors 
are large, the scatters in \bv\ and \ub\ are only of comparable sizes
to the observational errors, implying an intrinsically narrow sequence
in the age-metallicity parameter space.  

If we take the mean values of the derived ages and metallicities
(ignoring large errors), more metal-rich clusters 
%appear younger, indicating a physically plausible AMR. 
appear younger. 
There also seems to be a \ub\ (metallicity) bimodality with at least two
peaks at ${\rm [Fe/H]}\approx -$1 and 0. 
Zepf \& Ashman (1993) and Rejkuba (2002) also found two peaks but at
slightly different positions.

If this is true, the data seem to support the galaxy evolution 
scenario in which globular clusters started to form very early and 
subsequently at later redshifts. This seems consistent, 
albeit only qualitatively, with the prediction of semi-analytic models, 
where elliptical galaxies are results of numerous merger events. 
Harris \& Harris (2001) have reached a similar conclusion on this galaxy 
based on their halo giant star data.

The uncertainties in the age and metallicity estimates are 
currently too large to decide whether there indeed is a clear AMR.
It is very puzzling why there are so few metal-poor GCs
but so many super-metal-rich GCs compared to the GC systems in other galaxies. 
Does this make sense when viewed from the chemical evolution theory? 
If we are somehow underestimating reddening by 0.1~mag, the metallicity
distribution of the NGC\,5128 GCs would become similar to those of other
GC systems. Moreover, the observational errors are systematically larger for
redder GCs, and thus the age and metallicity estimates
of redder GCs are less certain.
Besides, our conclusions hinge on the slope of the GC 
locus in $UBV$ space. This slope is sensitive to the $U$-band photometric
colour term which needs to be measured carefully.
Another source of uncertainty is that population synthesis models 
have been rigorously tested only for metal-poor populations (mainly MWGCs) 
but not for metal-rich ones. Thus, the AMR the data seem to indicate 
may not be pristine. 
If we accordingly admit large uncertainties in our analysis, a single age 
(anywhere between 5 and 13~Gyr) hypothesis should not to be ruled out
yet (see Fig.~7). 

Despite the fact that we cannot conclude with confidence at this moment, 
it is encouraging that the data shows a narrow sequence in age and 
metallicity. 
A factor of two better-quality (signal-to-noise) data would lead to a much
stronger conclusion. Achieving reliable population models may require
substantially more work. Meanwhile, a simple but useful test can 
be made by acquiring spectra of some of the bright 
red clusters, e.g., those in the \ub = 0.5 -- 0.6 bin in Table~2. 
We should measure their spectroscopic metallicities and ages 
(e.g., using Balmer line indices) to validate/calibrate the population 
models. For example, if such analyses indicate that these GCs are 
substantially younger than typical old GCs, then it would confirm the 
AMR that we found. On the other hand,
if their spectra suggest old ages that are comparable to the ages of
the metal-poor GCs, a small-range of formation epoch for the GCs
would be more likely\footnote{After this work was completed, we learned that 
Peng et al. (2003) have measured spectroscopic line indices for the 
brightest of these clusters and found that the metal-rich GCs are indeed
young on average with the mean age of 4\,Gyr.}. 
Obviously, different galaxy evolution scenarios 
will be favoured depending on the result.

Another important test is to see whether these GCs show strong UV fluxes.
Yi (2003a) pointed out that the GCs that are old enough to develop a
large number of blue horizontal branch stars must exhibit a measurable 
UV flux, as shown in some of the MWGCs. Far-UV observations using HST or
GALEX will effectively differentiate old GCs from young ones.

One might point out that, although $U$-band data are useful in the
metallicity determination, it is substantially more difficult to obtain and
calibrate, and so the errors are large.  
It is true. NGC\,5128 is perhaps one of only a few galaxies to which 
we can apply
this technique using 4-m telescopes. Getting deep, precise $U$-band 
photometry for GCs in Virgo cluster galaxies, for example, would be 
much more difficult.  Nevertheless, it is an effective tool for deriving
metallicities unlike other dominantly popular colours such as \bv\ and \vi, 
which suffer from the age-metallicity degeneracy. 

We believe that this technique, when tested and calibrated sufficiently, 
can be applied to extragalactic populations of various 
properties to systematically study the star formation history in galaxies. 
It is possible to detect globular 
clusters in $U$-band even at the Virgo cluster distance 
if 8-m class telescopes are used. In addition,
UV observations (using HST and GALEX) are highly recommended because 
they could serve as effective age indicators for old metal-rich populations 
where the current $UBV$ two-colour technique becomes ineffective.

\section*{Acknowledgments}

We thank the anonymous referee for a number of constructive criticisms
and suggestions. 
We thank Joseph Silk, Ignacio Ferreras, Roger Davies, Robert Zinn, 
Pierre Demarque, Richard Larson, for valuable comments.
Stimulating discussions with Ray Sharples and Jean Brodie also shed
great lights on this study.
This research has been supported by PPARC Theoretical Cosmology Rolling Grant
PPA/G/O/2001/00016 and the Lockey Bequest Funds (S. K. Yi), 
and Glasstone Fellowship (S. Yoon). E. W. Peng acknowledges support from the 
National Science Foundation through grant AST 00-98566.

%\newpage

\end{document}

\end{document}